\documentclass{PoS}
\usepackage{amssymb,amsmath}

\newcommand{\vek}[1] {\mathbf{#1}}

\title{Correlations in Multi-Parton Interactions}

\ShortTitle{Correlations in Multiple-Parton Interactions}

\author{\speaker{Tomas Kasemets}\\%
        Deutsches Elektronen-Synchrotron DESY, 22603 Hamburg, Germany\\
        E-mail: \email{tomas.kasemets@desy.de}}

\abstract{In double parton interactions, the two hard proceses are correlated via double parton densities. We examine the double Drell-Yan
process and the impact of such correlations on the differential cross section. In particular correlations between the spins of the interacting quarks can induce correlations between the
decay planes of the vector bosons. We investigate upper limits on
spin correlations following from positivity of the double parton densities.}

\FullConference{Sixth International Conference on Quarks and Nuclear Physics\\
		 April 16-20, 2012\\
		 Ecole Polytechnique, Palaiseau,  Paris}

\begin{document}

\section{Introduction}
The increase in the density of partons with the energy of a hadron collider increases the amount of interactions where more than one parton in each hadron participates in hard interactions. Multi-parton interactions are therefore expected to play a larger role at the LHC than ever before. They contain information about the proton structure not available from single-parton interactions, and they can be important backgrounds to other signal processes, such as Higgs production \cite{DelFabbro:1999tf}. Multi-parton interactions have an experimental history stretching from the ISR \cite{Akesson:1987ab} to the LHC \cite{Dobson:2011}, and have long been modeled in event generators \cite{Buckley:2011ms}. However, the treatment of multi-parton interactions from the point of perturbative QCD has only recently started to accelerate. 

The hard interactions are connected via parton distributions and the partons originating from the same hadron can be correlated. This leads to interferences and correlation effects not present in single-parton scatterings. These effects are often neglected in studies of multiple hard scatterings, but the validity of such approaches still has to be investigated. A suitable scene for such a study is set by the double Drell-Yan process \cite{Mekhfi:1983az,Gaunt:2010pi,Manohar:2012jr}, where two quark/anti-quark pairs annihilate into two vector bosons ($\gamma^*$, $Z$, $W^\pm$) which decay leptonically. This process has the advantage of being theoretically clean and well understood in the single-parton scattering case. We calculate the differential cross section, taking initial parton correlations into account, and derive constraints on the size of the spin correlations following from positivity of double parton densities.
\section{Double Parton Interactions}
\begin{figure}
\centering
\includegraphics[width=0.7\textwidth]{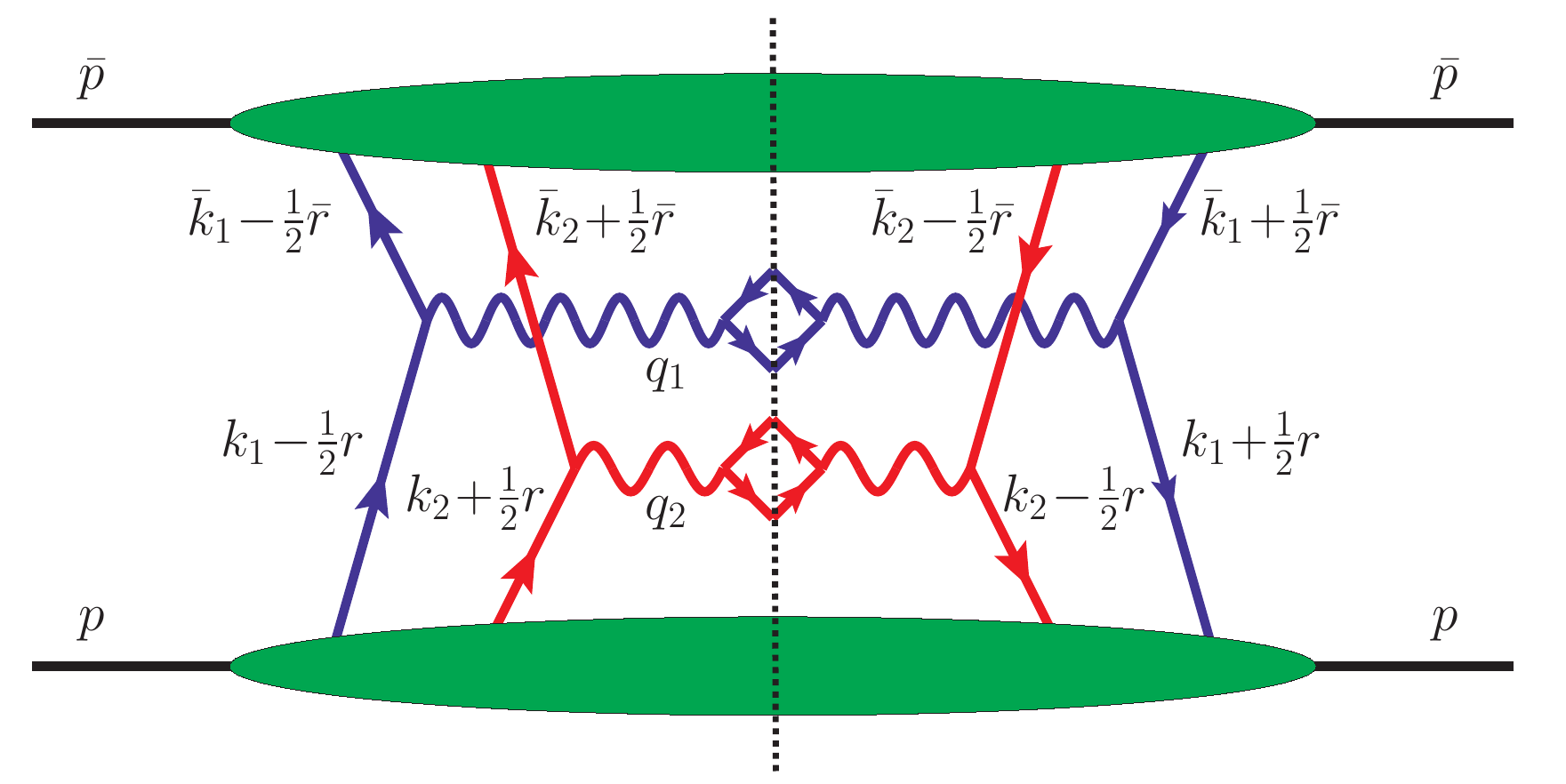}
\caption{\label{figures:DYmomentum}The double Drell-Yan process where two quarks in the right moving proton interact with two anti-quarks from the left moving proton. $q_i$ is the momentum of the vector boson from interaction i. $k_1$ and $k_2$ ($\bar{k}_1$ and $\bar{k}_2$) are average momenta carried by the partons from the right (left) moving proton taking part in hard interaction 1 and 2 respectively and $r$ the momentum difference.}
\end{figure}

When two partons in a proton interact, it is only the sum of momenta and quantum numbers which have to be the same in the amplitude and its conjugate. This allows for a momentum difference, $r$, between a parton in the amplitude and its partner in the conjugate amplitude, see figure \ref{figures:DYmomentum}. This difference has to be balanced by the parton in the other interaction. Similarly the color and flavor of the partons can be matched inside each hard collision or as an interference effect between them. There can even be fermion number interference between quarks and anti-quarks but we do not include that in our calculations. Further the spin of the interacting partons can be correlated, much in the same way as in single-parton collisions with polarized protons \cite{Boer:1999mm}. The correlations between the initial state partons are described by double parton distributions, DPDs. 
\subsection{Double Parton Distributions}
In transverse position space the DPDs depend on the collinear momentum fractions $x_1$ and $x_2$ of the two partons taking part in hard collisions, on their differences in transverse positions between the amplitude and its conjugate $\vek{z}_1$ and $\vek{z}_2$ and the transverse distance between the hard vertices $\vek{y}$. $\vek{z}_1$ ($\vek{z}_2$) is the Fourier conjugate of the average transverse momentum $\vek{k}_1$ ($\vek{k}_2$) and $\vek{y}$ of the momentum difference $\vek{r}$, see figure \ref{figures:DYmomentum}. The DPDs describing the different quark polarizations are labeled by $q$ for unpolarized, $\Delta q$ for longitudinally polarized and $\delta q$ for transversely polarized quarks \cite{Diehl:2011yj}.

For unpolarized and longitudinally polarized quarks the possible combinations are
\begin{equation}\label{eq:DPD1}\begin{aligned}
F_{qq} &= f_{qq}(x_1,x_2,\vek{z}_1,\vek{z}_2,\vek{y}), &
F_{q \Delta q} &= g_{q \Delta q} (x_1,x_2,\vek{z}_1,\vek{z}_2,\vek{y}), \\
F_{\Delta q \Delta q} &= f_{\Delta q \Delta q}(x_1,x_2,\vek{z}_1,\vek{z}_2,\vek{y}), &
F_{\Delta q q} &= g_{\Delta q q}(x_1,x_2,\vek{z}_1,\vek{z}_2,\vek{y}),
\end{aligned}\end{equation}
where $f$'s are scalar- and $g$'s are pseudo scalar-functions. Distributions with transversely polarized quarks in one of the two hard interactions carry an open index corresponding to the transverse spin vector, and have to be expanded in a basis spanning the transverse plane
\begin{equation}\begin{aligned}
F_{\Delta q \delta q}^i &= M\left(y^if_{\Delta q \delta q} + \tilde{y}^ig_{\Delta q \delta q}\right),&
F_{q \delta q}^i  &= M\left(\tilde{y}^if_{q\delta q} + y^ig_{q\delta q}\right) 
\end{aligned}\end{equation}
and similarly with the subscripts interchanged. $M$ is the proton mass and $\tilde{y}^i=y^j\epsilon^{ij}$, is a transverse vector orthogonal to $y^i$. Transversely polarized quarks in both interactions give two transverse indices and we need a tensor basis 
\begin{equation}\begin{split}
\label{eq:DPD3}
F_{\delta q \delta q}^{ij} &= \delta^{ij}f_{\delta q \delta q} + \left( 2y^i y^j - y^2\delta^{ij}\right)M^2f_{\delta q\delta q}^{\,t} + \left(y^i\tilde{y}^j + \tilde{y}^iy^j\right)M^2g_{\delta q\delta q}^{\,s} \\
&\hspace{0.3cm}+ \left(y^i\tilde{y}^j-\tilde{y}^iy^j\right)M^2g_{\delta q\delta q}^{\,a}.
\end{split}\end{equation}
Taking color interference into account gives one color singlet and one color interference distribution for each flavor combination and thus doubles the number of DPDs. When the flavors of the quarks are different there are flavor square and interference distributions. The DPDs for the left moving proton will be denoted by a bar, $\bar{f}_{qq}$, which should not be confused with the bar appearing over subscripts (indicating anti-particles).
\begin{figure}
\centering
\includegraphics[width=0.7\textwidth]{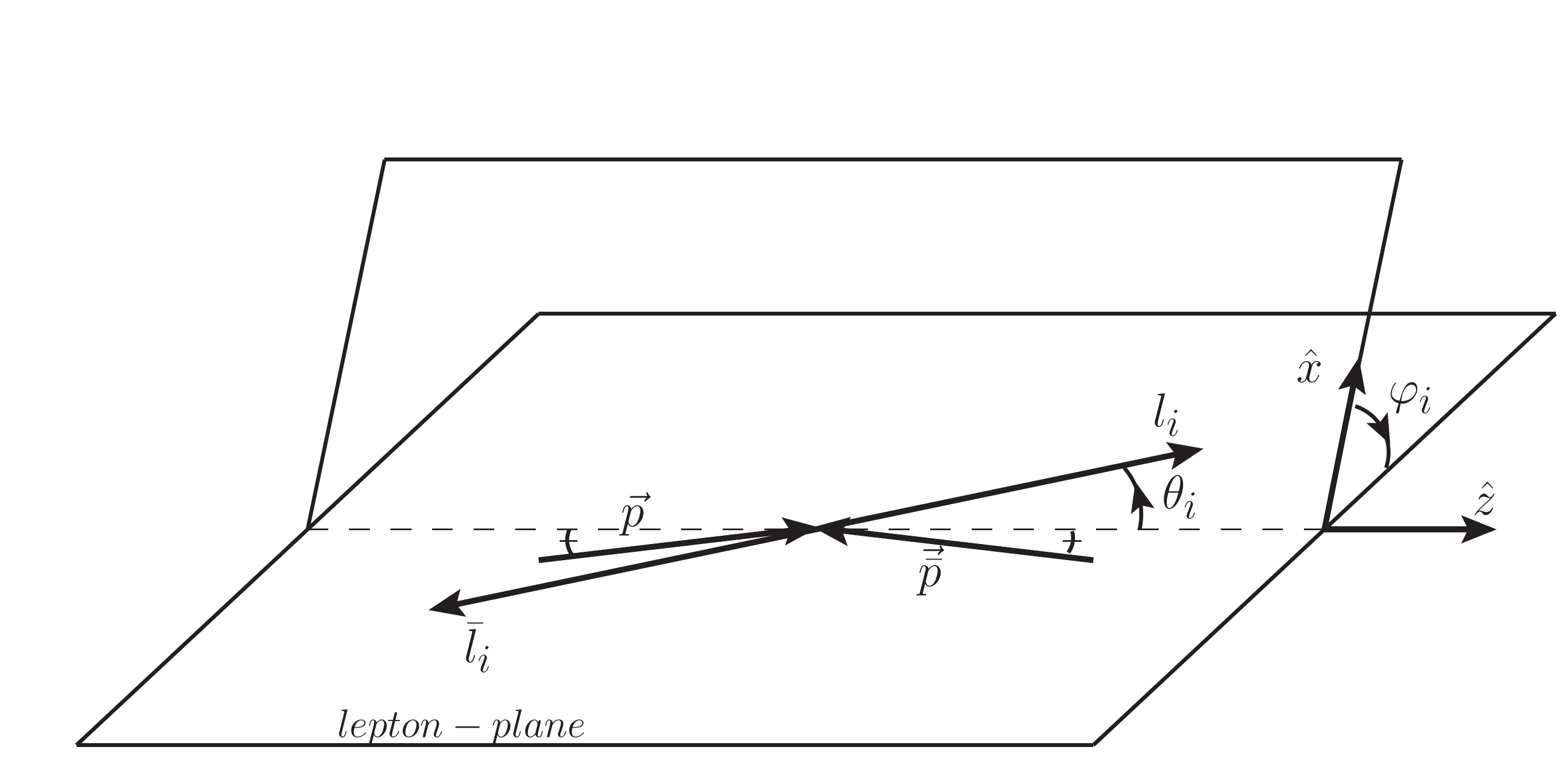}
\caption{\label{figures:angles}Reference frame for each interaction is the rest frame of the vector boson. $\theta_i$ is the polar and $\varphi_i$ the azimuthal angle of the lepton. $\vec{l}_i$ ($\vec{\bar{l}}_i$) is the three momentum of the lepton (anti-lepton) from interaction $i$.}
\end{figure}
\section{Cross Section}
The partonic cross sections are calculated to leading order for unpolarized, longitudinally and transversely polarized quarks and the results are combined with the DPDs describing the corresponding parton densities.

The cross section formula, assuming transverse momentum dependent factorization, derived in \cite{Diehl:2011yj}, can be expressed as
\begin{equation}\begin{split}
\label{eq:cross}
&\frac{d\sigma}{\prod_{i=1}^2dx_id\bar{x}_id^2\vek{q}_id\Omega_i}=\frac{1}{C}\sum_{q_1q_2\bar{q}_3\bar{q}_4}\sum_{a_1a_2\bar{a}_3\bar{a}_4}\frac{d\hat{\sigma}_{a_1\bar{a}_3}}{d\Omega_1}\frac{d\hat{\sigma}_{a_2\bar{a}_4}}{d\Omega_2}\int\frac{d^2\vek{z}_1}{(2\pi)^2}e^{-i
\vek{z}_1\vek{q}_1}\int\frac{d^2\vek{z}_2}{(2\pi)^2}e^{-i
\vek{z}_2\vek{q}_2}\\
&\hspace{1.2cm}\times\int d^2\vek{y}\left[F_{a_1a_2}\bar{F}_{\bar{a}_3\bar{a}_4}+F_{a_1\bar{a}_4}\bar{F}_{\bar{a}_3a_2}+F_{\bar{a}_3a_2}\bar{F}_{a_1\bar{a}_4}+F_{\bar{a}_3\bar{a}_4}\bar{F}_{a_1a_2}\right]+\left\{\text{\small flavor interference}\right\}.
\end{split}\end{equation}
$C$ equals 2 when the final states of the two hard interactions are identical and 1 otherwise. In the first sum $q_1$ ($q_2$) labels the flavor of the quark and $\bar{q}_3$ ($\bar{q}_4$) of the anti-quark entering the first (second) hard scattering. $a_i$ and $\bar{a}_i$ label the different combinations of unpolarized ($q_i$) and longitudinally ($\Delta q_i$) and transversely ($\delta q_i$) polarized quarks and anti-quarks. The color square and interference distributions enter the cross section with equal weights and we can therefore keep color square/interference labels implicit. $d\hat{\sigma}_i/d\Omega_i$ is the partonic cross section differential in the direction of the outgoing lepton.

To describe the final state we use angles defined in the rest frames of the vector bosons, with $\hat{z}$-axes defined as the direction bisecting the angle between $\vec{\bar{p}}$ and $-\vec{\bar{p}}$ ($\vec{p}$ and $\vec{\bar{p}}$ are three momenta of the protons), see figure \ref{figures:angles}. The $\hat{x}$-axis is an arbitrary transverse direction which we for definiteness choose to point towards the center of the LHC ring. The differences between the axes defined for the two vector bosons are of order $\vek{q}_i/Q_i$ and can be neglected.

 We will display the final results only for the cross sections integrated over transverse boson momenta explicitly, not including flavor interference, and refer to \cite{Diehl:2012me} for the cross sections depending on transverse boson momenta.

Integrating over the transverse momenta of the bosons yields collinear double parton distributions  \cite{Mekhfi:1983az}. After integration the DPDs depend on $x_1$, $x_2$ and only one transverse vector, $\vek{y}$, and therefore no pseudo-scalar functions can contribute. Due to time reversal symmetry, functions with one longitudinal and one transversely polarized quark vanish, for example  $f_{\Delta q \delta q}=0$. We are then left with six collinear double parton distributions for each combination of quark flavors. For unpolarized and longitudinally polarized quarks the cross section is
\begin{equation}\begin{split}
\frac{d\sigma^{(0)}}{dx_i\bar{x}_id\Omega_i} & = \frac{1}{C}\sum_{q_1q_2\bar{q}_3\bar{q}_4} \bigg\{ \left[K_{q_1\bar{q}_3}(1+\cos^2\theta_1)+K_{q_1\bar{q}_3}'\cos\theta_1\right]\left[K_{q_2\bar{q}_4}(1+\cos^2\theta_2)+K_{q_2\bar{q}_4}'\cos\theta_2\right]\hspace{-0.5cm} \\
& \hspace{1.8cm}\times \int d^2\vek{y} \hspace{0.1cm}\left[f_{q_1q_2}\bar{f}_{\bar{q}_3\bar{q}_4}+f_{\Delta q_1\Delta q_2}\bar{f}_{\Delta\bar{q}_3\Delta \bar{q}_4} + \text{perm.}\right]\\
&\hspace{0.9cm}+ \left[K_{q_1\Delta \bar{q}_3}(1+\cos^2\theta_1)+K_{q_1\Delta \bar{q}_3}'\cos\theta_1\right]\left[K_{q_2\Delta \bar{q}_4}(1+\cos^2\theta_2)+K_{q_2\Delta \bar{q}_4}'\cos\theta_2\right] \\
& \hspace{1.8cm}\times \int d^2\vek{y} \hspace{0.1cm}\left[f_{q_1q_2}\bar{f}_{\Delta \bar{q}_3\Delta \bar{q}_4}+f_{\Delta q_1\Delta q_2}\bar{f}_{\bar{q}_3\bar{q}_4}+ \text{perm.}\right]\bigg\}.
\end{split}\end{equation}
 $K_{q \bar{q}}$, $K_{q \bar{q}}'$, $K_{q\Delta \bar{q}}$ and $K_{q\Delta \bar{q}}'$ are $Q^2$ dependent coupling factors for different polarizations, defined in \cite{Diehl:2012me}. $\Omega_i$'s are solid angles and $\theta_i$'s polar angles of the leptons. 'perm' stands for permutations of the quark/anti-quark subscripts in the DPDs. 
For $W^\pm$ the coupling factors are polarization independent and including partonic spin correlations therefore only changes the rate, while for the neutral current the longitudinal polarization changes both the rate and the angular distribution of the process.

The part of the cross section with one interaction involving transversely polarized quarks vanish upon integration over the transverse boson momenta while the cross section with both interactions containing transversely polarized quarks is non-zero for the neutral current,
\begin{equation}\begin{split}
\frac{d\sigma^{(2)}}{dx_id\bar{x}_id\Omega_i} =&\, \frac{1}{C}\sum_{q_1q_2} \sin^2\theta_1\sin^2\theta_2 \int d^2\vek{y} \bigg\{\Big[(K_{\delta q_1 \delta q_1}K_{\delta q_2 \delta q_2}-K_{\delta q_1 \delta q_1}'K_{\delta q_2 \delta q_2}')\cos2(\varphi_1-\varphi_2)\hspace{-1cm}\\
&\hspace{-1cm}-(K_{\delta q_1 \delta q_1}'K_{\delta q_2 \delta q_2}-K_{\delta q_1 \delta q_1}K_{\delta q_2 \delta q_2}')\sin2(\varphi_1-\varphi_2) \Big]2\left[f_{\delta q_1\delta q_2}\bar{f}_{\delta\bar{q}_1\delta\bar{q}_2}+\text{perm.}\right]\bigg\}.
\end{split}\end{equation}
$K_{\delta q\delta \bar{q}}$ and $K'_{\delta q\delta \bar{q}}$ are $Q^2$ dependent coupling factors for two transversely polarized quarks and the $\varphi_i$'s are the azimuthal angles of the final state leptons. 
For $W$ bosons the coupling factors are zero for transversely polarized quarks, since the $W$ only couples to left handed quarks. The cross section for the neutral current depend on the azimuthal angle between the two outgoing leptons ($\varphi_1-\varphi_2$). This dependence originates in the correlations between the spin of the initial state quarks.
\section{Positivity Bounds}
The double parton distributions of different polarizations can be organized in a positive semi-definite spin density matrix. The positivity can then be used to derive upper limits on the polarized distributions, similarly to what was done for generalized parton distributions in \cite{Diehl:2005jf} and for transverse momentum dependent distributions in \cite{Bacchetta:1999kz}.

The projection operators $\Gamma_{++}$ ($\Gamma_{--}$) project out quarks with positive (negative) helicities while $\Gamma_{-+}$ and $\Gamma_{+-}$ give helicity interferences. These can be expressed in terms of operators projecting out un- ($\Gamma_q$), longitudinal- ($\Gamma_{\Delta q}$) and transversely- ($\Gamma_{\delta q}$) polarized quarks
\begin{equation}\begin{aligned}
\Gamma_{++} &= \frac{\gamma^+}{2}(1+\gamma_5) = \Gamma_q+\Gamma_{\Delta q}, &
\Gamma_{-+} &= -\frac{i\sigma^{+1}}{2}(1+\gamma_5) = \Gamma_{\delta q}^1-i\Gamma_{\delta q}^2, \\
\Gamma_{--} &= \frac{\gamma^+}{2}(1-\gamma_5) = \Gamma_q-\Gamma_{\Delta q}, &
\Gamma_{+-} &= \frac{i\sigma^{+1}}{2}(1-\gamma_5) = \Gamma_{\delta q}^1+ i\Gamma_{\delta q}^2.
\end{aligned}\end{equation}
 The DPDs can then be organized as a matrix in the light-cone helicity basis 
\begin{equation}\begin{split}
\hspace{-0.5cm}
\left( \begin{matrix}\vspace{0.1cm} f_{qq}+f_{\Delta q\Delta q} & -i|y|Mf_{\delta qq} & -i|y|Mf_{q\delta q} & 2y^2M^2f_{\delta q\delta q}^{\,t}\\\vspace{0.1cm}
i|y|M f_{\delta q q} & f_{qq}-f_{\Delta q\Delta q} & 2f_{\delta q \delta q} & -i|y|Mf_{q \delta q} \\
\vspace{0.1cm}i|y|Mf_{q\delta q} & 2f_{\delta q \delta q} & f_{qq}-f_{\Delta q\Delta q} & -i|y|Mf_{\delta qq} \\\vspace{0.1cm}
2y^2M^2f_{\delta q \delta q}^{\,t} & i|y|Mf_{q\delta q} &i|y|Mf_{\delta qq} & f_{qq}+f_{\Delta q \Delta q}
\end{matrix}\right)
\end{split}\end{equation}
where the rows (columns) correspond to $++,-+,+-,--$ helicities of the two quarks in the amplitude (conjugate amplitude). The non-interference parton distributions can be interpreted as the probability of finding quarks having specific helicities inside a proton. The helicity matrix is therefore positive semi-definite and calculating the eigenvalues we obtain necessary and sufficient positivity conditions
\vspace{-0.2cm}
\begin{equation}\begin{split}
f_{qq} \geq |f_{\delta q\delta q}-y^2M^2f_{\delta q\delta q}^{\,t}|
\end{split}\end{equation}
\begin{equation}\begin{split}
\Big(f_{qq}\pm (f_{\delta q\delta q}-y^2M^2f_{\delta q\delta q}^{\,t})\Big)^2-\Big(f_{\Delta q\Delta q}\mp(f_{\delta q \delta q}+y^2M^2f_{\delta q\delta q}^{\,t})\Big)^2
\geq y^2M^2\Big(f_{\delta q q}\pm f_{q\delta q}\Big)^2.
\end{split}\end{equation}
This implies the somewhat more transparent but weaker inequalities 
\begin{equation}\begin{aligned}
f_{qq}+f_{\Delta q \Delta q} & \geq 2y^2M^2f^{\,t}_{\delta q\delta q} & 
f_{qq}-f_{\Delta q \Delta q} & \geq 2f_{\delta q \delta q}.
\end{aligned}\end{equation}
\section{Summary}
Spin correlations as well as color and flavor interference result in a large number of double parton distributions. Many of these distributions contribute to similar terms in the double Drell-Yan cross section. Longitudinal polarization changes the overall rate for the charged bosons, while for the neutral bosons the angular distribution is also affected. The spin vectors of transversely polarized quarks lead to a dependence on the azimuthal angle between the two outgoing leptons. 
Similar features as those appearing in the double Drell-Yan cross section are expected to be present also in other types of processes, such as double dijet production, with larger cross sections but a dramatic increase of complexity due to their color structure.

We used the probability interpretation of double parton distributions to derive constraints on the polarized parton distributions. Similar constraints have proven very useful for single parton distributions.
\section{Acknowledgments}
I am grateful to Markus Diehl for collaboration and helpful discussions.

\end{document}